\documentclass[twocolumn,superscriptaddress,preprintnumbers,amsmath,amsfonts,amssymb]{revtex4-2}
\usepackage{graphicx}
\usepackage{bm}
\usepackage[bookmarks=false,colorlinks=true,citecolor=blue,linkcolor=blue,urlcolor=blue]{hyperref}

\begin{document}

\preprint{Journal reference: \href{https://doi.org/10.1103/PhysRevB.108.195108}{Phys.\ Rev.\ B \textbf{108}, 195108 (2023)}}

\title{Linear dichroic x-ray absorption response of Ti-Ti dimers along the $c$ axis
in Ti$_2$O$_3$ upon Mg substitution}

\author{M.~Okawa} 
\affiliation{Department of Applied Physics, Waseda University, Tokyo 169-8555, Japan}

\author{D.~Takegami} 
\affiliation{Max Planck Institute for Chemical Physics of Solids, 01187 Dresden, Germany}

\author{D.~S.~Christovam} 
\affiliation{Max Planck Institute for Chemical Physics of Solids, 01187 Dresden, Germany}

\author{M.~Ferreira-Carvalho} 
\affiliation{Max Planck Institute for Chemical Physics of Solids, 01187 Dresden, Germany}
\affiliation{Institute of Physics II, University of Cologne, 50937 Cologne, Germany}

\author{C.-Y.~Kuo} 
\affiliation{Max Planck Institute for Chemical Physics of Solids, 01187 Dresden, Germany}
\affiliation{Department of Electrophysics, National Yang Ming Chiao Tung University, Hsinchu 30010, Taiwan}
\affiliation{National Synchrotron Radiation Research Center (NSRRC), Hsinchu 30076, Taiwan}

\author{C.~T.~Chen} 
\affiliation{National Synchrotron Radiation Research Center (NSRRC), Hsinchu 30076, Taiwan}

\author{T.~Miyoshino} 
\affiliation{Department of Applied Physics, Waseda University, Tokyo 169-8555, Japan}

\author{K.~Takasu} 
\affiliation{Graduate School of Science and Engineering, Kagoshima University, Kagoshima 890-0065, Japan}

\author{T.~Okuda} 
\affiliation{Graduate School of Science and Engineering, Kagoshima University, Kagoshima 890-0065, Japan}

\author{C.~F.~Chang}  
\affiliation{Max Planck Institute for Chemical Physics of Solids, 01187 Dresden, Germany}

\author{L.~H.~Tjeng} 
\affiliation{Max Planck Institute for Chemical Physics of Solids, 01187 Dresden, Germany}

\author{T.~Mizokawa}
\affiliation{Department of Applied Physics, Waseda University, Tokyo 169-8555, Japan}

\date{\today}

\begin{abstract}
Corundum oxide Ti$_2$O$_3$ shows the metal-insulator transition around 400--600 K accompanying
the nearest Ti$^{3+}$-Ti$^{3+}$ bond ($a_{1g}a_{1g}$ singlet state) formation  along the $c$ axis.
In order to clarify the hole-doping effect for the $a_{1g}a_{1g}$ singlet bond  in Ti$_2$O$_3$, we investigated Ti $3d$
orbital anisotropy between corundum-type Ti$_2$O$_3$ and ilmenite-type MgTiO$_3$ using linear dichroism of soft x-ray
absorption spectroscopy of the Ti $L_{2,3}$ edge.
From the linear dichroic spectral weight in Mg$_y$Ti$_{2-y}$O$_3$, we confirmed that the $a_{1g}a_{1g}$
state is dominant not only in $y=0.01$ (almost Ti$_2$O$_3$), but also in $y = 0.29$, indicating that
the Ti-Ti bond survives against a certain level of hole doping.
In $y=0.63$ corresponding to 46\% hole doping per Ti, the $3d$ orbital symmetry changes from $a_{1g}$ to $e_g^{\pi}$.
\end{abstract}

\maketitle

\section{Introduction}
\begin{figure}[b]
\centering
\includegraphics[width=8.6cm]{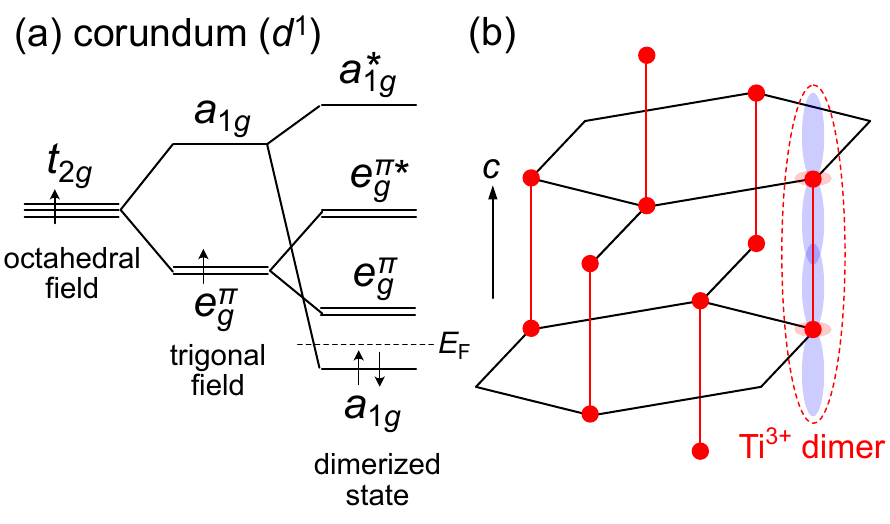}
\caption{
(a) Energy diagram of the $d$ orbitals in the corundum-type $d^1$ insulators with formation of $d^1$-$d^1$
singlet bond \cite{Castellani1978}.
(b) The Ti-Ti bond formation between honeycomb layers along the $c$ axis in Ti$_2$O$_3$.
The shaded areas are a schematic of Ti $3d$ $a_{1g}$ orbitals.
}
\label{schematic}
\end{figure}

Orbital degrees of freedom frequently play an essential role in the electronic properties of transition-metal oxides \cite{Khomskii2014}.
The metal-insulator transitions in rutile-type VO$_2$, for example, are governed by the bond formation between the V 3$d$ orbitals in
the edge sharing VO$_6$ octahedra \cite{Haverkort2005}.
There are also a variety of orbitally assisted bond formations in spinel and hollandite systems which were studied theoretically 
\cite{Khomskii2005,Khomskii2021} and experimentally \cite{Isobe2002,Schmidt2004,Zhou2005,Isobe2006,Ishige2011,Yamaguchi2022}.
Yet, in corundum-type V$_2$O$_3$, the bond formation surprisingly does not occur between the V ions in the face sharing pairs of VO$_6$
octahedra along the $c$-axis \cite{Park2000}.
In contrast, Ti-Ti molecular orbital formation does occur in Ti$_2$O$_3$:
The Ti-Ti distance of the pair is gradually shortened in going from 600 K to 400 K, and Ti$_2$O$_3$ undergoes a metal-to-insulator transition
(Fig.\ \ref{schematic}) \cite{Morin1959, Zandt1968, Robinson1974, Rice1977}.
As shown in Figs.\ \ref{schematic}(a) and (b), it has theoretically been shown that the Ti 3$d$ $a_{1g}$ orbitals form the molecular orbitals which build up the insulating state
\cite{Zeiger1975, Mattheiss1996, Tanaka2004, Poteryaev2004, Eyert2005, Castellani1978}.
The theoretical predictions have been followed by the experimental confirmations by soft x-ray absorption spectroscopy
(XAS) \cite{Sato2006,Chang2018} and photoemission spectroscopy (PES) \cite{Chang2018}.
In addition to the corundum Ti$_2$O$_3$ with $d^1$ honeycomb layers coupled by the face-sharing TiO$_6$ pairs, MgVO$_3$,
which is a novel ilmenite system with $d^1$ configuration, has been found to form V-V dimers in the $d^1$ honeycomb lattice below 500 K \cite{Yamamoto2022}.
It is also known that layered Ti trihalides such as TiCl$_3$ and TiBr$_3$ harbor the $d^1$ honeycomb lattice with Ti-Ti dimerization
\cite{Ogawa1960,Pei2020}.

Very recently, the impact of Mg substitution for Ti in the Mg$_y$Ti$_{2-y}$O$_3$ system, which is also known as Mg$_{1-x}$Ti$_{1+x}$O$_3$
with $y=1-x$ (hereafter, we use the notation with $y$), has been studied by Takasu \textit{et al.} \cite{Takasu2023}.
While Ti$_2$O$_3$ is the well studied corundum system, MgTiO$_3$ is an ilmenite system with Ti$^{4+}$ ($d^0$) configuration.
Therefore, Mg$_y$Ti$_{2-y}$O$_3$ provides a unique opportunity to study evolution from the face-sharing TiO$_6$ pairs in the $d^1$
corundum system to the $d^0$/$d^1$ mixed valence honeycomb lattice in the ilmenite.
It is expected that the Ti $3d$ $t_{2g}$ orbitals play essential roles to control their electronic properties.
It has been revealed that the shortened Ti-Ti bonds survive against the hole doping by the Mg$^{2+}$ substitution for Ti$^{3+}$. 
Here, the interesting question arises whether the Ti 3$d$ $a_{1g}$ orbitals are still occupied in the hole-doped system to stabilize
the Ti-Ti molecular orbitals.

As shown in the previous XAS studies \cite{Sato2006,Chang2018}, the particular linear dichroism (LD) of the Ti $2p$ spectrum between
the polarization vector ($\bm{E}$) perpendicular to and parallel to the $c$ axis played a vital role for clarifying the $a_{1g}$ occupation
in the Ti$_2$O$_3$ system.
In the present work, we report LD-XAS of Mg$_{y}$Ti$_{2-y}$O$_3$ with $y=0.01, 0.29, 0.63,$ and 1.00 to clarify the hole-doping effect
on the $a_{1g} a_{1g}$ singlet bond.
Based on the experimental results, we discuss the interplay between orbital symmetry and hole doping.

\section{Experiment and Calculations}
Single crystals of Mg$_{y}$Ti$_{2-y}$O$_3$ with $y=0.01$, 0.29, 0.63, and 1.00 were grown using the floating-zone method,
and the compositions $y$ were determined by the energy-dispersive x-ray spectroscopy, whose details were described in the literature
by Takasu {\it et al}.\ \cite{Takasu2023}.
The Ti ions in the octahedral coordination are substituted by the Mg ions.
The single crystals were mounted on the sample holders {\it ex situ} after orientation by the single crystal x-ray diffraction
using a Rigaku R-AXIS RAPID II diffractometer.
LD-XAS measurements with the total electron yield method were performed at the NSRRC-MPI TPS 45A1 Submicron
Soft X-ray Spectroscopy Beamline \cite{Tsai2019} at Taiwan Photon Source, National Synchrotron Radiation Research Center (NSRRC).
Clean sample surfaces were obtained by cleaving the crystals {\it in situ}, with the $c$-axis nearly in-plane of the cleaved surface.
LD-XAS data were acquired using a 98\% horizontal linear polarized soft x-ray beam, at a geometry close to normal incidence.
The LD spectra were obtained by first setting the sample stage to measure with the $c$-axis aligned parallel to the polarization direction,
and then rotating the stage in-plane 90$^\circ$ to obtain the geometry with the $c$-axis aligned perpendicular to the polarization.
The maximum dichroism at the chosen directions was verified to confirm the correct orientations.
The sample temperature was 300 K for all measurements, and an overall energy resolution was $\sim$0.25 eV.

In order to extract information about the orbital occupation from the XAS spectra, we have made use of simulations utilizing
the well-proven single-site TiO$_6$ and double-site Ti$_2$O$_9$ cluster-model calculations performed by
Sato {\it et al}.\ \cite{Sato2006} and by Chang {\it et al}.\ \cite{Chang2018}, respectively.
The method includes the full atomic multiplet theory and the local effects of the solid. It accounts for the intra-atomic Ti $3d$-Ti $3d$
and Ti $2p$-Ti $3d$ Coulomb interactions, the atomic Ti $2p$ and Ti $3d$ spin-orbit couplings, the O $2p$-Ti $3d$ hybridization,
and the proper local crystal-ﬁeld parameters
\footnote{
The parameters used in the TiO$_6$ cluster, in eV, are the following \cite{Chang2018}:
$U_{dd} = 4.0$, $U_{pd} = 5.5$, $\Delta=6.5$, $10Dq_\text{ionic}=0.85$,
$V_{e_g}^\sigma = 3.5$, $V_{e_g}^\pi = 1.2$, $V_{a_{1g}}^\pi = 0.9$. 
$\Delta_\text{trg}=-0.16$ and $0.16$ for the $e_g^\pi$ initial state and $a_{1g}$ initial state calculations respectively. The Ti$_2$O$_9$ clusters use the same parameters, with the addition of the Ti-Ti hopping integral V$_{dd\sigma}=0.9$.
}.

\section{Results and Discussion}

\begin{figure}[t]
\centering
\includegraphics[width=8cm]{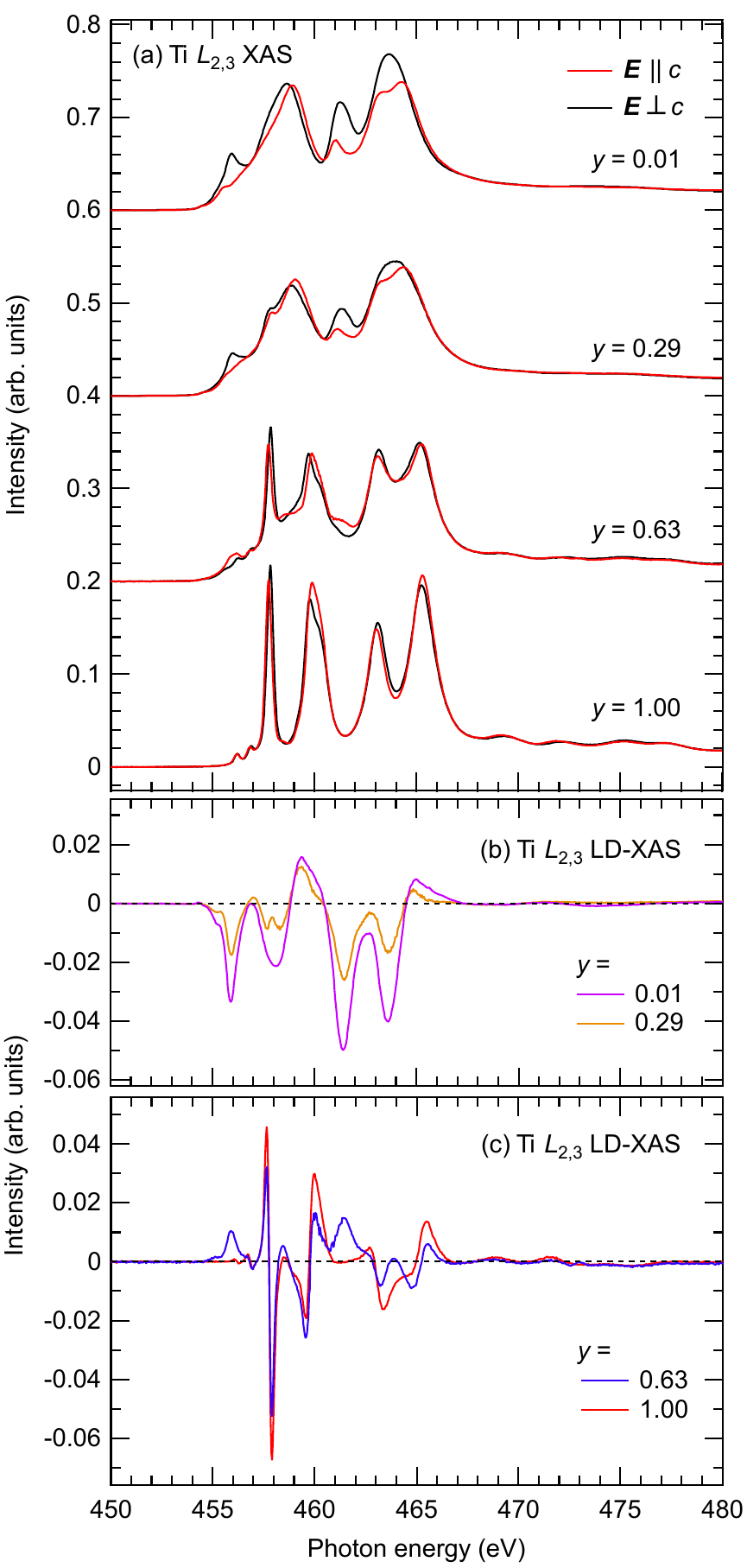}
\caption{
(a) Ti $L_{2,3}$ XAS of Mg$_y$Ti${_{2-y}}$O$_3$ for $y = 0.01$, 0.29, 0.63, and 1.00. 
The XAS intensities obtained with the linear-polarized direction of the incident beam ($\bm{E}$) parallel to
and perpendicular to the $c$-axis correspond to red ($\bm{E} \parallel c$) and black ($\bm{E} \perp c$) curves, respectively.
LD-XAS spectra obtained by the difference between the $\bm{E} \parallel c$ and $\bm{E} \perp c$ geometries were shown in
(b) for $y=0.01$ and 0.29, and in (c) for $y=0.83$ and 1.00.
Dashed lines correspond to zero LD intensity.
}
\label{Ti_XAS}
\end{figure}

Figure \ref{Ti_XAS}(a) shows the linear polarization dependence of the Ti $L_{2,3}$ edges for each $y$ composition.
The energy regions of 454--461 eV and 461--467 eV can be assigned to the $L_3$ ($2p_{3/2} \rightarrow 3d$) and
$L_2$ ($2p_{1/2} \rightarrow 3d$) absorption edges, respectively, although there is considerable overlap or mixing between the two edges \cite{deGroot1990a,deGroot1990b}.
In the compositions of $y = 0.63$ and 1.00, the sharp peaks at 458 eV and 460 eV are related (but not equal) to the octahedral crystal field
splitting at the Ti$^{4+}$ sites \cite{deGroot1990a}.
Ti $L_{2,3}$ LD were obtained as spectral differences between $\bm{E} \parallel c$ and $\bm{E} \perp c$ geometries for each $y$ as shown
in Fig.\ \ref{Ti_XAS}(b).
In $y = 0.01$, which is almost Ti$_2$O$_3$ composition, the LD-XAS spectrum is consistent with the previous reports of Ti$_2$O$_3$ at
room temperature \cite{Sato2006,Chang2018}.
The LD features in $y=0.01$ also appeared in the LD spectrum of $y = 0.29$ with a small Ti$^{4+}$ component, which can be seen at 458 eV.
In contrast to $y=0.01$ and 0.29, Ti $L_{2,3}$ XAS spectra in heavily Mg-substituted $y=0.63$ and MgTiO$_3$ ($y=1.00$) have very different
line shapes.
They are dominated by the Ti$^{4+}$ component as we will show below.

\begin{figure}[t]
\centering
\includegraphics[width=8.6cm]{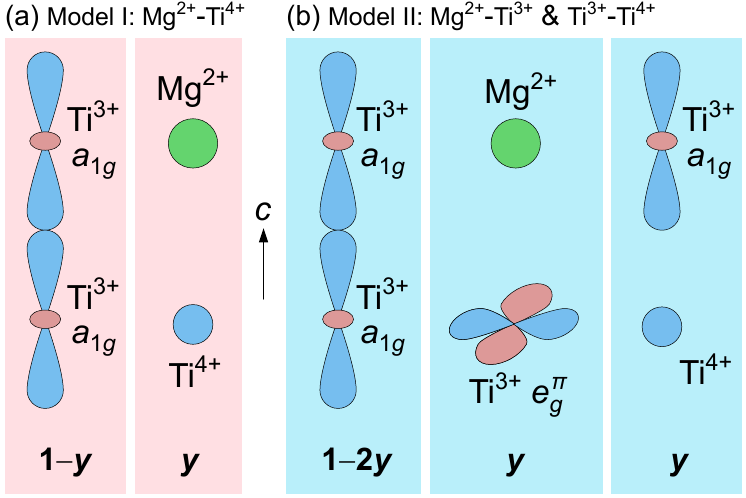}
\caption{
(a) Model I, corresponding to a Ti$^{3+}$-Ti$^{3+}$ pair with occupied $a_{1g}$-$a_{1g}$ orbitals and Mg$^{2+}$-Ti$^{4+}$ pairs with empty Mg $3s$ and Ti $3d$ orbitals.
Their relative populations are $1-y$ and $y$ respectively.
(b) Model II, corresponding to a Ti$^{3+}$-Ti$^{3+}$ pair with occupied $a_{1g}$-$a_{1g}$ orbitals, Mg$^{2+}$-Ti$^{3+}$ pairs with empty Mg $3s$ and occupied Ti $3d$
$e_g^{\pi}$ orbitals, and Ti$^{3+}$-Ti$^{4+}$ pairs with occupied Ti $3d$ $a_{1g}$ and empty Ti $3d$ orbitals.
Their relative populations are $1-2y$, $y$, and $y$, respectively.
}
\label{model}
\end{figure}

\begin{figure}[t]
\centering
\includegraphics[width=8.6cm]{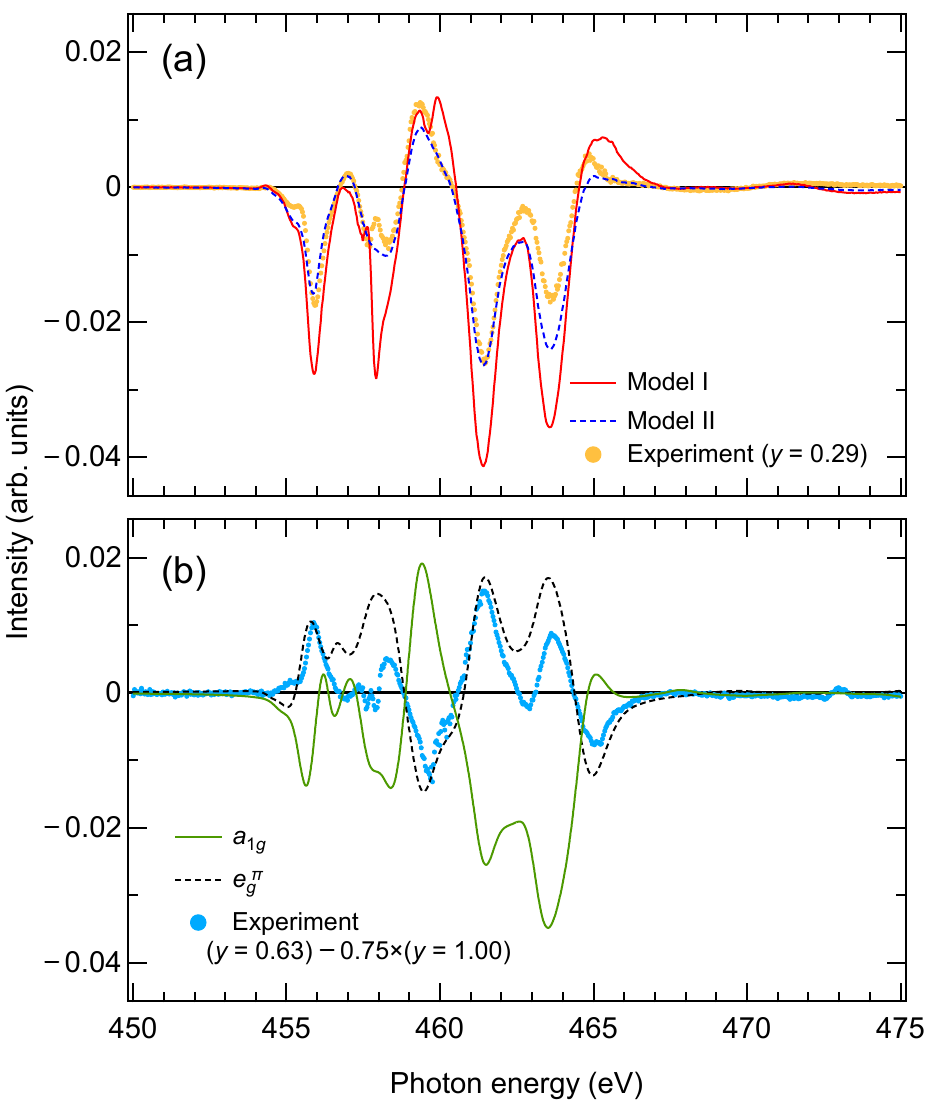}
\caption{
(a) Calculated LD spectra for $y = 0.29$ using the relative populations of the Ti sites corresponding to models I (red solid curve) and II (blue broken curve), as defined in Figs.\ \ref{model}(a) and (b), respectively and further elaborated in the text,
together with the experimental LD (yellow dots).
For model I, the calculations for the double-site Ti$_2^{3+}$O$_9$ cluster are adapted
from the work performed by Chang \textit{et al.} \cite{Chang2018}.
For model II, the calculations for the single-site Ti$^{3+}$O$_6$ cluster are reproduced
with permission from Sato \textit{et al.} \cite{Sato2006}.
(b) LD spectra of the Ti$^{3+}$ ions (blue dots), obtained from the difference between the LD intensities of $I_\text{LD}(y = 0.63)$
and $0.75 \times I_\text{LD}(y = 1.00)$, compared to the TiO$_6$ cluster-model calculations assuming the $a_{1g}$ (green solid curve) or
$e_g^\pi$ (black broken curve) orbital occupied.
Here we finely aligned the $I_\text{LD}(y = 1.00)$ by shifting it 8 meV to prevent derivative artifacts from the Ti$^{4+}$ contributions.
}
\label{LD_calc}
\end{figure}

In analyzing the LD spectrum of the $y=0.29$ composition, we first tried to construct it from the LD spectra of the $y=0.01$ and $y=1.00$
compositions with weights that follow from the $y=0.29$ value.
This first model is illustrated in Fig.\ \ref{model}(a) where the $y=0.01$ is represented by Ti$^{3+}$-Ti$^{3+}$ pairs with occupied
$a_{1g}$-$a_{1g}$ 3$d$ orbitals and the $y=1.00$ by Mg$^{2+}$-Ti$^{4+}$ pairs with empty Mg 3$s$ and Ti 3$d$ orbitals, respectively. 
We then obtained a sum of the LD of the $y=0.01$ multiplied by $2(1-0.29)/(2-0.29) = 0.83$ and the LD of the $y=1.00$ multiplied by
$0.29/(2-0.29) = 0.17$, where we note that the factors $1/(2-0.29)$ is to account for the fact that the spectra in Fig.\ \ref{Ti_XAS}(a) with
different Ti contents were normalized to the integrated intensity.
The result is shown in Fig.\ \ref{LD_calc}(a) and clearly does not match the measured LD of the $y=0.29$.
The latter shows mainly the same LD of the $y=0.01$ but reduced by about 50\% and very little of the LD of the $y=1.00$.

This discrepancy let us to conclude that the substitution of Ti by Mg does not create Mg$^{2+}$-Ti$^{4+}$ pairs but rather
Mg$^{2+}$-Ti$^{3+}$ pairs with the Ti$^{4+}$ being located somewhere else and thus forming Ti$^{3+}$-Ti$^{4+}$ pairs.
This conclusion is consistent with the findings of a recent photoemission study which revealed that the holes introduced by the Mg substitution
are in the Ti-Ti pairs \cite{Miyoshino2023}.
We thus arrive at the second model as depicted in Fig.\ \ref{model}(b): the $y=0.29$ material is composed of Mg$^{2+}$-Ti$^{3+}$ pairs
with the empty Mg 3$s$ orbital and the singly occupied Ti 3$d$ orbital (of the $e_g^{\pi}$ type as we will show later),
Ti$^{3+}$-Ti$^{4+}$ pairs with the singly occupied 3$d$ orbital (of the  $a_{1g}$ type) and the fully unoccupied 3$d$ orbitals, respectively,
and Ti$^{3+}$-Ti$^{3+}$ pairs with $a_{1g}$-$a_{1g}$.
The relative weights of the components are also indicated, namely, $y$, $y$, and $1-2y$, respectively.
The LD spectra of the Ti$^{3+}$ ions in the Mg$^{2+}$-Ti$^{3+}$ and Ti$^{3+}$-Ti$^{4+}$ pairs can be estimated from the earlier
TiO$_6$ cluster calculations \cite{Sato2006,Chang2018} as also shown in Fig.\ \ref{LD_calc}(b).
Making the sum of the LD spectra from an $e_g^{\pi}$ Ti$^{3+}$ and an $a_{1g}$ Ti$^{3+}$ cluster plus the LD from the $y=0.01$ with
0.29, 0.29, and $2(1-2 \times 0.29)$ weights, respectively, followed by the $1/(2-0.29)$ multiplication, we obtain a result that is very close to
the experimental LD of $y=0.29$ as displayed in Fig.\ \ref{LD_calc}(a).
We note that assuming the Ti$^{3+}$ in the Mg$^{2+}$-Ti$^{3+}$ and Ti$^{3+}$-Ti$^{4+}$ pairs are all $e_g^{\pi}$ or all $a_{1g}$
gives less satisfactory results.

Focusing now on the $y=0.63$ composition, we observe that the sharpest features of the LD spectrum are given by those of the $y=1.00$,
i.e., by the Ti$^{4+}$ in the Mg$^{2+}$-Ti$^{4+}$ pairs.
The LD spectra of the $y=0.63$ and $y=1.00$ are, however, not identical and the difference between them contains information about
the orbital state of the Ti$^{3+}$ in the Mg$^{2+}$-Ti$^{3+}$ and Ti$^{3+}$-Ti$^{4+}$ pairs.
Fig.\ \ref{LD_calc}(b) shows this difference spectrum by the dotted curve.
We can observe spectral features that matches well with those of the $e_g^{\pi}$ in the TiO$_6$ cluster, e.g., the peaks at 455 eV, 459.5 eV,
461.5 eV, 462.5 eV, 463.5 eV, and 465 eV.
This supports our conjecture from the $y=0.29$ composition in that the Ti$^{3+}$ in the Mg$^{2+}$-Ti$^{3+}$ pair is of the $e_g^{\pi}$ type.
At the same time, we now also see that in this high Mg content $y=0.63$ composition the Ti$^{3+}$ of the Ti$^{3+}$-Ti$^{4+}$ pairs are
partly converted from $a_{1g}$ to $e_g^{\pi}$.

\begin{figure}[t]
\centering
\includegraphics[width=7.75cm]{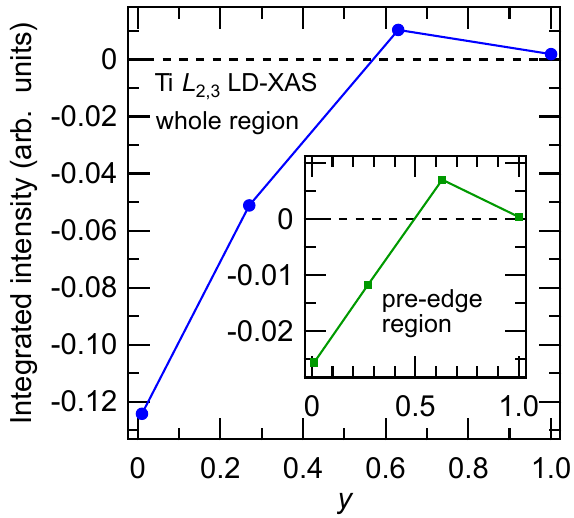}
\caption{
Whole integrated intensity (450--470 eV) of Ti $L_{2,3}$ LD-XAS as a function of $y$ in Mg$_{y}$Ti$_{2-y}$O$_3$.
Inset: The integrated intensity around the pre-edge region (455--457 eV).
The negative values in $y=0.01$ and 0.29 [i.e., $I_\text{LD}(\bm{E} \parallel c) < I_\text{LD}(\bm{E} \perp c)$] mean
dominating the Ti-Ti singlet bonds.
The sign change from negative to positive at $y=0.63$ mainly corresponds to the peaks at 456 eV and 462 eV shown in Fig.\ 2(c).
}
\label{LD_area}
\end{figure}

To confirm our findings, we also analyzed the integrated intensities of the spectra and their polarization dependence since the use of sum rules
gives direct information about the orbital polarization of the ground state \cite{Csiszar2005,HaverkortPhD}.
Figure \ref{LD_area} shows the integrated intensity of the Ti $L_{2,3}$ LD-XAS spectra for the whole energy region (450--470 eV) and pre-edge
region (455--457 eV).
In $y \leq 0.29$ samples, integrated LD-XAS intensities are negative, i.e., $I(\bm{E} \parallel c) < I(\bm{E} \perp c)$.
The $a_{1g}$ orbital spreads along the $c$ axis, while the $e_g^{\pi}$ orbitals spread in the $a$-$b$ plane,
and thus the negative LD-XAS intensity can be expected by the relatively dominant $a_{1g}$ orbital occupancy,
suggesting that the Ti-Ti bonds along the $c$-axis are still robust in $y=0.29$.
On the other hand, in $y = 1.00$ (MgTiO$_3$), the integrated Ti $L_{2,3}$ LD-XAS intensity is almost zero.
This is consistent with the Ti$^{4+}$ $d^0$ configuration where all the $t_{2g}$ and $e_g$ orbitals are unoccupied.
As for $y=0.63$, the integrated value of Ti $L_{2,3}$ LD-XAS is positive, i.e., $I(\bm{E} \parallel c) > I(\bm{E} \perp c)$.
The positive integrated value is mainly  due to the positive LD signals at 456 eV (Ti $L_3$ pre-edge region)
and at 462 eV (Ti $L_2$ pre-edge region), which are not observed for $y = 1.00$ in Fig.\ \ref{Ti_XAS}(b). 
As shown in the inset of Fig.\ \ref{LD_area}, the LD signal of the Ti $L_3$ pre-edge region changes from negative to positive in going from
$y=0.29$ to $y=0.63$. 
The sign change of the LD signal indicates that symmetry of the unoccupied Ti 3$d$ orbitals changes between $y=0.29$ and
$y=0.63$.
In $y=0.63$, the $a_{1g}$ orbital is less occupied and the $e_g^\pi$ orbitals are more occupied than at $y=0.29$;
thus the $\bm{E} \parallel c$ intensity is enhanced by the unoccupied $a_{1g}$ state  [Fig.\ \ref{schematic}(a)].
Therefore, the Ti-Ti bonds along the $c$-axis tend to collapse at the hole-doping level for $y=0.63$.

The Ti $3d$ orbital occupation changes from $a_{1g}$ to $e_g^\pi$ with the Mg doping.
Since the $e_g^\pi$ orbitals are extended along the Ti-Ti bonds in the honeycomb lattice layers, the $e_g^\pi$ electrons can hop
in the honeycomb lattice in the $y=0.63$ system.
Without dimerization, the $y=0.63$ system is expected to be more metallic than the $y=0.29$ system.
On the other hand, while the $a_{1g}$ orbital stabilizes the Ti-Ti bond in the face sharing TiO$_6$ pairs, the $e_g^\pi$ orbital may favor
Ti-Ti dimers in the honeycomb lattice layer as observed in MgVO$_3$.
It would be interesting to study the $y=0.63$ system using extended x-ray absorption fine structure (EXAFS) or
pair distribution function (PDF) experiments in order to detect local Ti-Ti dimers in the honeycomb lattice.

\section{Conclusion}
We investigated the Mg$^{2+}$ substitution effect to the Ti-Ti bonds along the $c$-axis in Ti$_2$O$_3$
using LD-XAS measurements.

We found that the substitution of Ti by Mg does not lead to the formation of Mg$^{2+}$-Ti$^{4+}$ pairs as previously thought.
Instead, it creates Mg$^{2+}$-Ti$^{3+}$ pairs, with the Ti$^{4+}$ ions forming Ti$^{3+}$-Ti$^{4+}$ pairs elsewhere in the structure.
This conclusion was further supported by the observation that the holes introduced by the Mg substitution were found to be in the Ti-Ti pairs.
We established a model for the $y=0.29$ composition, which consists of Mg$^{2+}$-Ti$^{3+}$ pairs with $e_g^{\pi}$ orbital symmetry,
Ti$^{3+}$-Ti$^{4+}$ pairs with $a_{1g}$ orbital symmetry, and Ti$^{3+}$-Ti$^{3+}$ pairs with $a_{1g}$-$a_{1g}$ singlet bonds.
As the Mg content increased in the $y=0.63$ composition, we observed a change in the orbital symmetry of the Ti$^{3+}$ ions in
the Ti$^{3+}$-Ti$^{4+}$ pairs, with some converting from $a_{1g}$ to $e_g^{\pi}$.
The Ti-Ti bonds along the $c$-axis tend to collapse at this hole-doping level for $y=0.63$, indicating a significant impact of hole doping on
the electronic properties.
These findings in Mg$_{y}$Ti$_{2-y}$O$_3$ provide valuable insights into the electronic properties of transition-metal oxides and
their metal-insulator transitions.

\begin{acknowledgments}
This work was supported by JSPS KAKENHI Grant No.\ JP19H00659 and No.\ JP22H01172.
M.F.-C.\ benefited from the financial support of the German Research Foundation (DFG), Project No.\ 387555779. 
The authors used research equipment for Laue measurements (Rigaku R-AXIS RAPID II: Material Characterization Central Laboratory, Waseda University) shared in the MEXT Project for promoting public utilization of advanced research infrastructure (Program for supporting construction of core facilities), Grant No.\ JPMXS0440500022.
The authors also are thankful for support for the crystal sectioning by the Joint Research Center for Environmentally Conscious Technologies in Materials Science, ZAIKEN, Waseda University. The authors acknowledge support from the Max Planck-POSTECH-Hsinchu Center for Complex Phase Materials.
\end{acknowledgments}

%%%%%%%%%%%%%%%%%%%%%%%%%%%%%%%%%%%%%%%%%%%%%%%%%%%%%%%%%%%%%%%%%
%%%%%%%%%%%%%%%%%%%%%%%%%%%%%%%%%%%%%%%%%%%%%%%%%%%%%%%%%%%%%%%%

%\bibliographystyle{apsrev4-2}
%\bibliography{references}


\end{document}